\documentclass[12pt]{article}
\pdfoutput=1
\usepackage[DIV13]{typearea}
\usepackage{indentfirst}
\usepackage{amsmath}
\usepackage{mathrsfs}
\usepackage{amssymb}
\usepackage{bbm}
\usepackage{graphicx}
\usepackage{slashed}
\usepackage{multicol}
\usepackage[usenames,dvipsnames]{color}
\usepackage{cite}
\usepackage{cancel}
\usepackage[normalem]{ulem}
\usepackage{array}
\usepackage{multirow}

\RequirePackage[colorlinks=true,urlcolor=blue,anchorcolor=blue,citecolor=blue,filecolor=blue,
linkcolor=blue,menucolor=blue,linktocpage=true,pdfproducer=medialab]{hyperref}
\newcommand{\email}[1]{\href{mailto:#1}{\tt #1}}

\numberwithin{equation}{section}
\newcommand{\LL}{\mathscr{L}}

\def\cD{{\cal D}}
\def\cF{{\cal F}}
\def\cG{{\cal G}}
\def\cH{{\cal H}}

\def\cO{{\cal O}}

\def\cP{{\cal P}}
\def\cQ{{\cal Q}}
\def\cR{{\cal R}}
\def\cS{{\cal S}}
\def\cX{{\cal X}}

\def\Tr{{\rm Tr}}

\def\be{\begin{equation}}
\def\ee{\end{equation}}
\def\beq{\begin{equation}}
\def\eeq{\end{equation}}
\def\bc{\begin{center}}
\def\ec{\end{center}}
\def\bea{\begin{eqnarray}}
\def\eea{\end{eqnarray}}

\def\nn{\nonumber}
\newcommand{\TeV}{\;\text{TeV}}

\newcommand{\derp}{\partial}

\newcommand{\unity}{\mathbbm{1}}
\newcommand{\Bs}{B^*}
\newcommand{\Ws}{W^*}
\newcommand{\UH}{\mathbf{U}}
\newcommand{\SH}{\mathbf{\Sigma}}

\newcommand{\TL}{\mathbf{T}}
\newcommand{\VL}{\mathbf{V}}
\newcommand{\DL}{\mathbf{D}}
\newcommand{\BL}{\mathbf{B}}
\newcommand{\BLs}{\BL^*}
\newcommand{\WL}{\mathbf{W}}
\newcommand{\WLs}{\WL^*}

\newcommand{\gtt}{\mathfrak{g}}
\newcommand{\htt}{\mathfrak{h}}

\newcommand{\tr}{\Tr}

\newcommand{\cAt}{\widetilde{{\cal A}}}
\newcommand{\cBt}{\widetilde{{\cal B}}}
\newcommand{\ct}{\tilde{c}}

\newcommand{\TLt}{\widetilde{\mathbf{T}}}
\newcommand{\VLt}{\widetilde{\mathbf{V}}}
\newcommand{\BLt}{\widetilde{\mathbf{B}}}
\newcommand{\BLts}{\widetilde{\mathbf{B}}^*}
\newcommand{\WLt}{\widetilde{\mathbf{W}}}
\newcommand{\WLts}{\widetilde{\mathbf{W}}^*}

\newcommand{\SLt}{\widetilde{\mathbf{S}}}
\newcommand{\SLts}{\widetilde{\mathbf{S}}^*}

\newcommand{\vh}{\langle \varphi \rangle}
\newcommand{\alf}{\left[\dfrac{\varphi}{f}\right]}
\newcommand{\alfm}{\left[\dfrac{\varphi}{2f}\right]}
\newcommand{\alfd}{\left[\dfrac{2\varphi}{f}\right]}

\newcommand{\Dfb}{\mbox{$\raisebox{2mm}{\boldmath ${}^\leftrightarrow$}\hspace{-4mm} \DL^\mu$}}

\newcommand{\blue}[1]{\color{blue} #1 \color{black} }

\usepackage{catchfile}
\newcommand{\getenv}[2][]{%
  \CatchFileEdef{\temp}{"|kpsewhich --var-value #2"}{}%
  \if\relax\detokenize{#1}\relax\temp\else\let#1\temp\fi
}
\getenv[\USER]{USER}
\newcommand\redout{\bgroup\markoverwith
{\textcolor{red}{\rule[.5ex]{2pt}{0.4pt}}}\ULon}

\begin{document}
\begin{titlepage}
\vspace*{-1cm}
\phantom{hep-ph/***} 
{\flushleft
{\blue{DFPD-2015/TH/25}}
\hfill{\blue{FTUAM-15-35}}
\hfill{\blue{IFT-UAM/CSIC-15-113}}}

\vskip 1.5cm
\begin{center}
{\LARGE\bf Sigma Decomposition:}\\[3mm] 
{\LARGE\bf The CP-Odd Lagrangian}\\[3mm]
\vskip .3cm
\normalsize
\end{center}
\vskip 0.5  cm
\begin{center}
{\large I.M.~Hierro}~$^{a)}$,
{\large L.~Merlo}~$^{b)}$,
{\large and S.~Rigolin}~$^{a)}$
\\
\vskip .7cm
{\footnotesize
$^{a)}$~
Dipartimento di Fisica ``G.~Galilei'', Universit\`a di Padova and \\
INFN, Sezione di Padova, Via Marzolo~8, I-35131 Padua, Italy\\
$^{b)}$~
Instituto de F\'{\i}sica Te\'orica, IFT-UAM/CSIC,\\
Universidad Aut\'onoma de Madrid, Cantoblanco, 28049, Madrid, Spain\\
\vskip .1cm
\vskip .5cm
\begin{minipage}[l]{.9\textwidth}
\begin{center} 
\textit{E-mail:} 
\email{ignacio.hierro@pd.infn.it},
\email{luca.merlo@uam.es},
\email{stefano.rigolin@pd.infn.it}.
\end{center}
\end{minipage}
}
\end{center}

\vskip 1cm
\abstract{ 
In {\em Alonso et al., JHEP {\bf 12} (2014) 034}, the CP-even sector of the effective chiral Lagrangian for a generic 
composite Higgs model with a symmetric coset has been constructed, up to four momenta. In this paper, the CP-odd couplings are studied within the same context. If only the Standard Model bosonic sources of custodial symmetry breaking are considered, then at most six independent operators form a basis. One of them is the weak-$\theta$ term linked to non-perturbative sources of CP violation, while the others describe CP-odd perturbative couplings between the Standard Model gauge bosons and an Higgs-like scalar belonging to the Goldstone boson sector. The procedure is then applied to three distinct exemplifying frameworks: the original $SU(5)/SO(5)$ Georgi-Kaplan model, the minimal custodial-preserving $SO(5)/SO(4)$ model and the minimal $SU(3)/(SU(2)\times U(1))$ model, which intrinsically breaks custodial symmetry. Moreover, the projection of the high-energy electroweak effective theory to the low-energy chiral effective Lagrangian for a dynamical Higgs is performed, uncovering strong relations between the operator coefficients and pinpointing the differences with the elementary Higgs scenario.} 
\end{titlepage}
\setcounter{footnote}{0}

\tableofcontents

%
%
\newpage
\section{Introduction}

A plausible solution to the electroweak (EW) hierarchy problem is provided in the so-called composite Higgs (CH) scenario 
\cite{Kaplan:1983fs,Kaplan:1983sm,Banks:1984gj,Georgi:1984ef,Georgi:1984af,Dugan:1984hq}, where the Higgs particle arises 
as a pseudo Nambu-Goldstone Boson (GB) of a global symmetry breaking present at high energy. In this setup, the global 
symmetry group, denoted here $\cG$, is spontaneously broken by some strong dynamics mechanism to a subgroup, $\cH$, at a 
certain scale $\Lambda_s$ that for definiteness we assume around the TeV. Among the GBs arising from the breaking, three 
have to be identified with the would-be-longitudinal components of the Standard Model (SM) gauge bosons and one can be 
associated to a scalar field, playing the role of the Higgs field, $\varphi$. The characteristic scale of this global symmetry 
breaking, dubbed $f$, is related with the strong scale $\Lambda_s$ by the relation $\Lambda_s\le 4\pi f$ \cite{Manohar:1983md}. 
Due to the GBs shift symmetry, the Higgs develops a mass only at one-loop level due to an induced Coleman-Weinberg scalar potential: 
an elegant solution to the SM hierarchy problem is thus provided (for recent reviews see for example Refs.~\cite{Contino:2010rs, 
Panico:2015jxa}).

Smoking guns of specific CH models~\cite{Contino:2003ve,Agashe:2004rs,Contino:2006qr,Gripaios:2009pe} are exotic resonances 
with masses smaller than $1.5\TeV$~\cite{Matsedonskyi:2012ym,Pomarol:2012qf,Redi:2012ha,Marzocca:2012zn,Panico:2012uw, 
Pappadopulo:2013vca,Contino:2013gna,Matsedonskyi:2014lla}, that however have not been discovered at colliders yet. On the other 
hand, indirect studies on deviations from the SM predictions are viable strategies to test the presence of New Physics (NP). 
The most suitable tool in this case is the effective field theory approach, that model-independently provides a description 
of non-SM signals that could be seen at colliders.

When considering only the bosonic sector, the effective Lagrangian at the EW scale $v$ (defined by the $W$ mass $m_W=g\,v/2$) 
can be written in terms of the SM gauge bosons -- the longitudinal components, represented by the GB matrix
\beq
\UH(x)=e^{i\sigma_a \pi^a(x)/v}\, ,
\label{SMGBs}
\eeq
and the transverse ones $W_{\mu\nu}$ and $B_{\mu\nu}$ -- and of an isosinglet CP-even scalar $h$ representing the resonance 
discovered at LHC~\cite{Aad:2012tfa,Chatrchyan:2012ufa}. In the most general case, the couplings with $h$ can be described 
by a generic smooth function $\cF(h)$~\cite{Alonso:2012px}. The complete effective basis of pure-gauge and gauge-Higgs 
interactions\footnote{Fermionic operators have been 
discussed at different levels and with different aims in Refs.~\cite{Alonso:2012jc,Alonso:2012pz,Buchalla:2012qq,Buchalla:2013rka}.}
has been presented, respectively for the CP--even and CP--odd case, in Refs.~\cite{Alonso:2012px,Gavela:2014vra}, based on 
the Appelquist-Longhitano-Feruglio (ALF) basis~\cite{Appelquist:1980vg,Longhitano:1980iz,Longhitano:1980tm,Feruglio:1992wf, 
Appelquist:1993ka}, and following the spirit of Refs.~\cite{Grinstein:2007iv,Contino:2010mh,Azatov:2012bz}. 
This low-energy effective Lagrangian represents a fundamental tool for Higgs studies at collider as shown in 
Refs.~\cite{Brivio:2013pma,Brivio:2014pfa,Gavela:2014vra}, where the main focus was in disentangling an elementary Higgs 
from a composite one, by the analysis of its couplings. In the former case, the SM GBs are described together with the 
physical Higgs by the electroweak Higgs doublet, and this leads to correlations between certain observables and to the 
suppressions of the so-called anomalous couplings. On the other hand, in the composite case, the adimensionality of the 
GBs matrix $\UH(x)$, and the treatment of $\UH(x)$ and the Higgs field $h$ as independent objects translate into an 
additional decorrelations of observables and into the appearance of unsuppressed anomalous couplings.

The low-energy effective chiral Lagrangian described in Refs.~\cite{Alonso:2012px,Gavela:2014vra} can be useful to describe 
an extended class of BSM ``Higgs'' models, from more extreme technicolor-like ansatzs to intermediate situations such as CH 
models or  dilaton-like scalar frameworks. The distinct limits can be reached for different values of the GB scale $f$ and by 
fixing the value of the coefficient and choosing the specific $\cF(h)$ function associated to each effective operator. 
On the other hand, it is often interesting to acquire a top-bottom perspective and connect the low-energy effective chiral 
Lagrangian with specific classes of CH models. This has been worked out in detail in Ref.~\cite{Alonso:2014wta}, where the 
CP conserving high-energy effective chiral Lagrangian for a generic symmetric coset $\cG/\cH$ has been constructed, up to 
four momenta. Three representative examples have been then analysed: the original $SU(5)/SO(5)$ Georgi-Kaplan model, the minimal 
custodial-preserving $SO(5)/SO(4)$ model and the minimal $SU(3)/(SU(2)\times U(1))$ model, which intrinsically breaks 
custodial symmetry. The projection at low-energy of the effective Lagrangian for each of the aforementioned CH models 
is shown to match the chiral effective Lagrangian for a dynamical Higgs of Refs.~\cite{Alonso:2012px}, uncovering strong 
relations between the operator coefficients.

The aim of this paper is to complete the study performed in Ref.~\cite{Alonso:2014wta} by introducing the CP-odd effective 
chiral Lagrangian, up to four derivatives for a generic symmetric coset. The analysis is then detailed for the same three 
representative CH models as in Ref.~\cite{Alonso:2014wta}. This completes the tool necessary to study exotic gauge-Higgs 
couplings at colliders and at low-energy experiments, and to investigate the nature of the Higgs particle and the origin 
of the EWSB mechanism. 

The paper is organised as follows. Sect.~\ref{Sect:ExtendedALF} is devoted to recalling the low-energy effective chiral 
Lagrangian introduced in Refs.~\cite{Alonso:2012px,Gavela:2014vra}, focusing on the CP-odd couplings. Sect.~\ref{Sect:BasisGeneral} 
contains the high-energy effective chiral Lagrangian, describing the CP-odd interactions among SM gauge bosons and the 
GBs associated to the symmetric coset $\cG/\cH$. Only operators with at most four derivatives are retained in the Lagrangian. 
Furthermore, no source of custodial breaking besides the (gauge) SM one is considered. In Sects.~\ref{Sect:GKmodel}, 
\ref{Sect:MinimalSO5} and \ref{Sect:MinimalSU3}, the low-energy effective EW chiral Lagrangian is then derived from the 
high-energy one for the $SU(5)/SO(5)$, $SO(5)/SO(4)$ and $SU(3)/(SU(2)\times U(1))$ composite Higgs models. Finally in 
Sect.~\ref{Sect:Matching}, the connection with the low-energy chiral Lagrangian describing the linearly realised EW 
symmetry scenario (dubbed ``linear Lagrangian'' from now on), written in terms of the SM $SU(2)_L$-doublet Higgs, is 
also discussed. Conclusions are presented in Sect.~\ref{Sect:Conc}. 

%
%
\boldmath
\section{\cancel{CP} effective chiral Lagrangian at the EW scale}
\unboldmath
\label{Sect:ExtendedALF}

The Higgsless EW effective chiral Lagrangian described in Refs.~\cite{Appelquist:1980vg,Longhitano:1980iz,
Longhitano:1980tm,Feruglio:1992wf,Appelquist:1993ka} can be written in terms 
of the SM gauge bosons $W^a_\mu(x)$ and $B_\mu(x)$, the SM GBs matrix $\UH(x)$ and its covariant derivative: 
\beq
\DL_\mu \UH(x) \equiv \derp_\mu \UH(x) +ig\WL_{\mu}(x)\UH(x) - \dfrac{ig'}{2} B_\mu(x) \UH(x)\sigma_3 \,,
\eeq
where $\WL_\mu(x)\equiv W_{\mu}^a(x)\sigma_a/2$. To write the effective chiral operators it is convenient to introduce 
the following (pseudo-)scalar and vector chiral fields:
\beq
\TL(x)\equiv \UH(x) \sigma_3 \UH^\dagger(x)\,,\qquad\qquad
\VL_\mu(x)\equiv \left(\DL_\mu \UH(x)\right)\UH^\dagger(x)\,.
\label{oldchiral}
\eeq
Recalling the transformation property of $\UH(x)$ under a (global) $SU(2)_{L,R}$ transformation:
\beq
 \UH(x) \rightarrow L\, \UH(x) R^\dagger
\eeq
it follows that both $\TL(x)$ and $\VL(x)$ transform in the adjoint of $SU(2)_L$,
\beq
\TL(x)\rightarrow L\,\TL(x)L^\dagger\,,\qquad\qquad
\VL_\mu(x)\rightarrow L\,\VL_\mu(x)L^\dagger\,.
\eeq
These chiral fields (and their derivatives) together with the EW gauge bosons are the necessary building blocks to construct 
the (Higgsless) ALF basis. 

After the discovery of the new light scalar degree of freedom (aka Higgs particle) one is forced to extend the previous 
basis, including all possible couplings between the Higgs, taken in all generalities as a CP-even scalar singlet field $h$, 
the SM gauge bosons and the (pseudo-)scalar and vector chiral fields. The effective chiral Lagrangian at the EW scale, 
describing the gauge and gauge-Higgs interactions, up to four derivatives, has been derived in 
Ref.~\cite{Alonso:2012px,Gavela:2014vra}, for the CP-even and CP-odd sectors respectively. In the following, we will focus 
only on the CP-odd sector, and we list here the full set of operators~\cite{Gavela:2014vra} necessary to describe the 
CP-odd bosonic interactions, organising them by their number of derivatives and their custodial character: 
\begin{description}
\item[CP-odd operators with two derivatives]
\beq
\hspace{-4cm}
\begin{aligned}
&\underline{\text{Custodial preserving}}\hspace{2.5cm}
&&\underline{\text{Custodial breaking}} \\
&\hspace{1.8cm}- && \hspace{-0.2cm} \cS_{2D}= i\,\frac{v^2}{4}\,\text{Tr}\left(\TL\,\cD_\mu\VL^\mu\right)\\
&&
\end{aligned}
\label{ExtALFOp2}
\eeq
\item[CP-odd operators with four derivatives]
\beq
\hspace{-0.5cm}
\begin{aligned}
&\underline{\text{Custodial preserving}}\hspace{3.0cm} &&\underline{\text{Custodial breaking}}\\
& \cS_{BB^*}=-\dfrac{g^{\prime2}}{4} B^*_{\mu\nu}B^{\mu\nu} &&
\cS_{4}= g\text{Tr}\left(\WL^{\mu\nu}\VL_\mu\right)\text{Tr}\left(\TL\,\VL_\nu\right)\\
& \cS_{WW^*}=-\dfrac{g^2}{2}\tr(\WL^*_{\mu\nu}\WL^{\mu\nu}) &&
\cS_{5}= i\text{Tr}\left(\VL^\mu\,\VL^\nu\right)\text{Tr}\left(\TL\,\VL_\mu\right)\partial_{\nu}(h/v)\\
& &&
\cS_{6}= i\text{Tr}\left(\VL^\mu\,\VL_\mu\right)\text{Tr}\left(\TL\,\VL^\nu\right)\partial_{\nu}(h/v)\\
& \cS_{1}= 2g\,g'\,B^*_{\mu\nu}\text{Tr}\left(\TL \WL^{\mu\nu}\right) &&
\cS_{7}= g\, \text{Tr}\left(\TL\,\left[\WL^{\mu\nu},\VL_\mu\right]\right)\,\partial_{\nu}(h/v)\\
& \cS_{2}= 2\,i\,g'\,B^*_{\mu\nu}\,\text{Tr}\left(\TL\,\VL^\mu\right)\,\partial^{\nu}(h/v) &&
\cS_{8}= 2\,g^2\text{Tr}\left(\TL\,\WL^*_{\mu\nu}\right)\text{Tr}\left(\TL\,\WL_{\mu\nu}\right)\\
& \cS_{3}= 2\,i\,g\,\text{Tr}\left(\WL^*_{\mu\nu}\,\VL_\mu\right)\,\partial_{\nu}(h/v) &&
\cS_{9}= 2\,i\,g\text{Tr}\left(\WL^*_{\mu\nu}\,\TL\right)\text{Tr}\left(\TL\,\VL_\mu\right)\partial_{\nu}(h/v)\\
& &&
\cS_{10}= i\,\text{Tr}\left(\VL^\mu\,\cD^\nu\VL_\nu\right)\,\text{Tr}\left(\TL\,\VL_\mu\right)\\
& &&
\cS_{11}= i\,\text{Tr}\left(\TL\,\cD^\mu\VL_\mu\right)\,\text{Tr}\left(\VL^\nu\,\VL_\nu\right)\\
& &&
\cS_{12}= i\,\text{Tr}\left([\VL^\mu,\TL]\,\cD^\nu\VL_\nu\right)\,\partial_{\mu}(h/v)\\
& &&
\cS_{13}= i\,\text{Tr}\left(\TL\,\cD^\mu\VL_\mu\right)\,\partial^{\nu}\partial_{\nu}(h/v)\\
& &&
\cS_{14}= i\,\text{Tr}\left(\TL\,\cD^\mu\VL_\mu\right)\,\partial^{\nu}(h/v)\,\partial_{\nu}(h/v) \\ 
& &&
\cS_{15}= i\,\text{Tr}\left(\TL\,\VL^\mu\right)\,\left(\text{Tr}\left(\TL\,\VL^\nu\right)\right)^2\,\partial_{\mu}(h/v)\\
& &&
\cS_{16}= i\,\text{Tr}\left(\TL\,\cD^\mu\VL_\mu\right)\,\left(\text{Tr}\left(\TL\,\VL^\nu\right)\right)^2
\end{aligned}
\label{ExtALFOp4}
\eeq
\end{description}
where $X^*_{\mu\nu}\equiv \epsilon_{\mu\nu\rho\sigma} X^{\rho\sigma}$ for $X\equiv\{B,\,\WL\}$ and $\cD_\mu$ denotes the covariant derivative in the adjoint representation of $SU(2)_L$, i.e.
\beq
\cD_\mu\VL_\nu\equiv \derp_\mu \VL_\nu + i\,g\,\left[\WL_\mu,\VL_\nu\right]\,.
\eeq
The operators on the left column are custodial preserving, meaning that they do not introduce custodial breaking contributions distinct from the SM ones; instead, the operators on the right column encode tree-level custodial breaking sources beyond the SM (gauge) one. A common aspect of the operator in the last class is the presence of the scalar chiral field $\TL(x)$ not in association with the $B_{\mu\nu}$ field strength. Notice, however, that the presence of $\TL(x)$ inside an operator does automatically implies that it violates the custodial symmetry. For example, it is straightforward to verify that 
\beq
\epsilon_{\mu\nu\rho\sigma}\text{Tr}\left(\TL\,\VL^\mu\,\VL^\nu\right)\text{Tr}\left(\TL\,\VL^\rho\right)\derp^\sigma h=
\dfrac{2}{3}\epsilon_{\mu\nu\rho\sigma}\text{Tr}\left(\VL^\mu\,\VL^\nu\VL^\rho\right)\derp^\sigma h
\label{ExampleDoubleT}
\eeq
which shows that the presence of $\TL(x)$ on the l.h.s. does not imply a custodial breaking nature of the operator, as it 
can be rewritten as an obviously custodial preserving one. This happens only for the operator on the l.h.s of Eq.~(\ref{ExampleDoubleT}), due to the presence of two $\TL(x)$ in that specific combination. 

The CP--odd (Higgless) ALF basis can be obtained from Eqs.~(\ref{ExtALFOp2}) and (\ref{ExtALFOp4}), disregarding all 
the operators containing one or more derivatives of $h$. Specifically the ALF CP-odd basis consists of one operator with 
two derivatives, $\cS_{2D}$, plus one custodial preserving and five custodial breaking operators with four-derivatives, 
besides the topological ones.

The low-energy electroweak chiral Lagrangian describing the CP-odd gauge, gauge-Goldstone and the gauge-Higgs interactions 
can instead be written as:
\beq
\LL_\text{low,{\cancel{CP}}}=\LL^{p^2}_\text{low,{\cancel{CP}}}+\LL^{p^4}_\text{low,{\cancel{CP}}}\,,
\label{LALFgen}
\eeq
where $\LL^{p^2}_\text{low,{\cancel{CP}}}$ and $\LL^{p^4}_\text{low,{\cancel{CP}}}$ contain two and four-derivative 
operators, respectively,
\beq
\begin{aligned}
\LL^{p^2}_\text{low,{\cancel{CP}}} = & \,c_{2D}\cS_{2D}\cF_{2D}(h)\,,
\label{LALF}\\
\LL^{p^4}_\text{low,{\cancel{CP}}} = &\,\cS_{B\Bs} \cF_{B\Bs}(h) + \, \cS_{W\Ws}\cF_{W\Ws}(h) + 
          \sum_{i=1}^{16}c_i\,\cS_i\cF_i(h) \,,
\end{aligned}
\eeq
with the functions $\cF_i(h)$ encoding a generic dependence on $h$\footnote{Be aware that no derivatives of 
$h$ are included in $\cF_i(h)$.}. The careful reader should be warned about the slightly different notation used here with respect to that used in Refs.~\cite{Alonso:2012px,Gavela:2014vra} for the definition of the $\cF_i(h)$ functions. 
Here, the operators $\cP_i$ do not contain the $\cF_i(h)$ functions, which are instead left outside as multiplicative terms, and the only dependence on $h$ kept inside the operators $\cP_i$ is $\partial h$. Furthermore, in Refs.~\cite{Alonso:2012px,Gavela:2014vra}, it was made explicit the dependence on a 
parameter that qualifies the degree of non-linearity of the theory: this parameter is usually labelled $\xi$ and defined as
\beq
\xi\equiv (v/f)^2\,.
\label{xi}
\eeq
This parameter is meaningful only when the high-energy theory that leads to the low-energy effective chiral Lagrangian 
is specified. Here, to keep the discussion as general as possible, the $\xi$ weights are reabsorbed in the coefficients 
$c_i$ and in the functions $\cF_i(h)$. The role of $\xi$ will become clear in the following sections, once specific 
dynamical Higgs models will be considered. 

%
%
\boldmath
\section{\cancel{CP} effective chiral Lagrangian for symmetric cosets}
\unboldmath
\label{Sect:BasisGeneral}

In this section, the high-energy CP-odd effective Lagrangian will be constructed for a generic CH setup. The discussion will closely follow Ref.~\cite{Alonso:2014wta}, to which the reader is referred to for further details. Only few definitions will be recalled here and the notation will be fixed.

Consider a generic CH framework and denote with $\cG$ the global symmetry group, spontaneously broken by some strong dynamics mechanism at the scale $\Lambda_s$, to a subgroup $\cH$. The coset $\cG/\cH$ is assumed to be symmetric (see \cite{Low:2014nga}): this does not restrict the applicability of the results of this analysis as all the known existing CH models satisfy this condition. The minimum requisite on the choice of specific $\cG$ and $\cH$ is that $dim(\cG/\cH) \ge 4$, i.e. at least four GBs arise from the global symmetry breaking, such that three of them would be then identified with the longitudinal components of the SM gauge bosons and one with the light scalar resonance observed at LHC. No fermionic operators will be considered here, and only CP-odd ones will be retained among the set of bosonic operators, up to four derivatives. Moreover, we assume that the only sources of custodial breaking are the ones of the SM.

The GBs arising from the global spontaneous symmetry breaking $\cG\to\cH$ can be described by a field $\Omega(x)$ defined as~\cite{Coleman:1969sm,Callan:1969sn}
\beq
\Omega(x)\equiv e^{i\,\Xi(x)/2f}\,,
\label{DefOmega}
\eeq
transforming under the global groups $\cG$ and $\cH$ as 
\beq
\Omega(x)\rightarrow \gtt\, \Omega(x)\, \htt^{-1}(\Xi, \gtt)\,.
\label{TransOmega1}
\eeq
This expression defines the non-linear transformation of $\Xi(x)$: $\gtt$ represents a (global) element of $\cG$, while $\htt(\Xi, \gtt)$ a (local) element of $\cH$, which depends explicitly on $\gtt$ and on the Goldstone boson field $\Xi(x)$. Notice that, for sake of brevity, $\htt(\Xi, \gtt)$ will be simply denoted as  $\htt$ in the following.

The GB field matrix $\Xi(x)$ can be explicitly written in terms of the generators of the coset $\cG/\cH$, $X_{\hat{a}}$ (with $\hat{a}=1, \ldots,dim(\cG/\cH)$), as
\beq
\Xi(x)=\Xi^{\hat{a}}(x) \,X_{\hat{a}}\,.
\label{GBtot}
\eeq
The $X_{\hat{a}}$ generators together with the generators of the preserved subgroup $\cH$, $T_a$ (with $a=1,\ldots,dim(\cH)$), form an orthonormal basis of $\cG$.

The peculiarity of symmetric cosets is to admit an automorphism (usually dubbed ``grading'') that acts on the generators 
of $\cG$ as
\beq
\cR:\quad
\begin{cases}
T_a\rightarrow +T_a\\
X_{\hat{a}}\rightarrow -X_{\hat{a}}
\end{cases}
\label{grading}
\eeq
In a generic symmetric coset it is then possible to define a ``squared'' non-linear field $\SH(x)$:
\beq
\SH(x)\equiv \Omega(x)^2\,,
\label{Sigma} 
\eeq
transforming under $\cG$ as,
\beq
\SH(x)\rightarrow \gtt\, \SH(x)\, \gtt_{\cR}^{-1}\,.
\eeq
From Eq.~(\ref{grading}) one immediately recovers that the GB field matrices $\Omega(x)$ and $\SH(x)$ transform, under the 
discrete grading $\cR$ automorphism, respectively, as:
\beq
\Omega(x) \rightarrow \Omega(x)^{-1}\,, \qquad\qquad \SH(x) \rightarrow \SH(x)^{-1}\,.
\eeq
This shows the equivalence in using $\Omega(x)$ or $\SH(x)$ in the description of GBs degrees of freedom and its interactions, in frameworks with a symmetric coset $\cG/\cH$. Here, we will follow Ref.~\cite{Alonso:2014wta} and adopt the description of the CP-odd operators using the $\Sigma$--representation.

To introduce gauge interactions it is convenient to formally gauge the full group $\cG$. In the symmetric coset case, 
it is possible to define both the $\cG$ gauge fields $\SLt_\mu$, and the graded siblings $\SLt_{\mu}^{\cal R} \equiv 
\cR (\SLt_{\mu})$, transforming under $\cG$, respectively, as:
\beq
\SLt_{\mu} \rightarrow \gtt \, \SLt_{\mu} \, \gtt^{-1} - \frac{i}{g_S} \gtt (\partial_\mu \,\gtt^{-1}) \,,
\qquad\qquad
\SLt_{\mu}^{\cal R}\rightarrow \gtt_\cR\,\SLt_{\mu}^{\cal R}\,\gtt^{-1}_\cR-\frac{i}{g_S}\gtt_\cR(\partial_\mu\,\gtt^{-1}_\cR)\,,
\label{TransS2}
\eeq
with $g_S$ denoting the associated gauge coupling constant. The covariant derivative of the non-linear field $\SH(x)$ and 
the chiral vector field can then be defined as: 
\bea
\DL_\mu\SH &=&\derp_\mu \SH + i\,g_S (\SLt_\mu \SH - \SH\, \SLt^\cR_\mu) \,, \label{newcovdev} \\ 
\VLt_\mu &=& \left(\DL_\mu \SH \right) \SH^{-1} \,,
\label{newchiralg}
\eea
The following three $\cG$-covariant objects can thus be used as building blocks for the (gauged) effective chiral Lagrangian:
\beq
\VLt_\mu  \,, \qquad 
\SLt_{\mu\nu} \quad 
\text{and} \quad 
\SH\, \SLt_{\mu\nu}^\cR \, \SH^{-1}\,.
\eeq
The introduction of the graded vector chiral field $\VLt_\mu^\cR$ does not add any further independent structure, as indeed
\beq
\VLt_\mu^\cR \equiv \cR (\VLt_\mu) = \left(\DL_\mu \SH \right)^{-1} \SH 
\qquad  {\rm with} \qquad 
\SH\,\VLt_\mu^\cR\,\SH^{-1} = -\VLt_\mu \,.
\label{Vgrading}
\eeq

Under the hypothesis of absence of any custodial symmetry breaking source besides the SM ones, any operator containing the 
high-energy sibling of the scalar chiral field $\TL(x)$,
\beq
\TLt\equiv \SH\, Q_Y\, \SH^{-1},
\eeq 
with $Q_Y$ being the embedding in $\cG$ of the hypercharge generator, should not enter in the basis, except in one 
specific case, discussed later on, where the presence of two $\TLt$ gives rise to a custodial preserving operator. 

\subsection{Basis of independent operators}
Performing an expansion in momenta and keeping operators with at most four derivatives, one can write the following  
independent structures: 
\begin{description}
\item[4-momenta CP-odd operators with gauge field strength $\SLt_{\mu\nu}$]
\beq
\Tr\left(\SLts_{\mu \nu}\SLt^{\mu \nu}\right)\,,\qquad\qquad
\Tr\left(\SLts_{\mu\nu}\,\SH\,\SLt^{\mu\nu,R}\,\SH^{-1}\right)\,,
\label{GbasisS}
\eeq
The first operator resembles the usual $\theta$ term operator for QCD. The other contains 
gauge-GB and pure-gauge interactions. 
\item[4-momenta CP-odd operators without gauge field strength $\SLt_{\mu\nu}$]
\beq
\begin{gathered}
\epsilon_{\mu\nu\rho\sigma}\Tr\left(\TLt\left[\VLt^\mu\,,\VLt^\nu\right]\right)\Tr\left(\TLt\left[\VLt^\rho\,,\VLt^\sigma\right]\right) 
\,,\qquad\qquad
\epsilon_{\mu\nu\rho\sigma}\Tr\left(\VLt^\mu\,\VLt^\nu\VLt^\rho\,\VLt^\sigma\right)\,,
\end{gathered}
\label{GbasisVV}
\eeq
Other operators with traces of two $\VLt$'s are clearly vanishing due to antisymmetry. 
\end{description}
The operators listed in Eqs.~(\ref{GbasisS})--(\ref{GbasisVV}) represent the set of independent CP--odd structures that 
can be introduced in a generic symmetric coset. Other operators could be easily introduced, apparently giving rise 
to new independent structures, most notably:
\beq
\Tr\left(\SLts_{\mu\nu}\left[\VLt^\mu,\VLt^\nu\right]\right)\,, \quad 
\epsilon_{\mu\nu\rho\sigma}\Tr\left((\cD^\mu\VLt^\nu)\,(\cD^\rho\VLt^\sigma)\right)\,, \quad 
\epsilon_{\mu\nu\rho\sigma}\Tr\left((\cD^\mu\VLt^\nu)\,\VLt^\rho\,\VLt^\sigma\right)\,, \nn
\eeq
where the adjoint covariant derivative acting on $\VLt^\mu$ is defined as
 \beq
 \cD_\mu\VLt_\nu= \derp_\mu \VLt_\nu+i\, g_S\left[\SLt_\mu,\VLt_\nu\right]\,. \nn
 \eeq
However, making use of the Bianchi identity and of the well known properties of the commutator of the covariant derivatives, 
it is possible to show that all these operators do not introduce any independent structure besides the one listed in 
Eqs.~(\ref{GbasisS})--(\ref{GbasisVV}). It is also worth noticing that in specific $\cG/\cH$ realizations, some of the 
operators listed may not be independent. For example the operator with traces of four $\VLt^\mu$ appearing on the right hand side of Eq.~(\ref{GbasisVV}) is redundant in all the considered CH models even if it was not the case for its CP-even counterpart. Moreover, other operators containing even powers of $\TLt$ could be considered in the list above, but it is straightforward to show that they introduce custodial breaking sources beyond the SM ones, and therefore are not 
considered. \footnote{We thank A.~Wulzer for the discussions that led to include the operator on the right hand side of Eq.~(\ref{GbasisVV}) in the list of independent operators.}

\boldmath
\subsection{General EW effective Lagrangian for a symmetric $\cG/\cH$ coset}
\unboldmath

The operators listed in Eqs.~(\ref{GbasisS})--(\ref{GbasisVV}) have been obtained under the hypothesis of gauging the full group $\cG$. In most of the CH realisations, however, only the SM gauge group is gauged, while the group $\cG$ is global. In these cases, the generic gauge field $\SLt_\mu$ should contain only the components corresponding to the EW symmetry. 

Specifying the dependence on the EW gauge bosons does not lead to new operator structures in the sector made out exclusively of  $\VLt_\mu$ fields (see Eq.~(\ref{GbasisVV})), while all operators where the gauge field strength appears explicitly, such as those in Eq.~(\ref{GbasisS}), should be ``doubled'' by substituting 
$\SLt_\mu$ either with $\WLt_\mu$ or $\BLt_\mu$, defined by
\beq
\WLt_\mu\equiv W^a_\mu \,Q^a_L \qquad {\rm and} \qquad \BLt_\mu \equiv B_\mu \,Q_Y\,,
\label{WBSU(5)}
\eeq
where $Q^a_L$ and $Q_Y$  denote the embedding in $\cG$ of the $SU(2)_L\times U(1)_Y$ generators. It follows that a larger number of independent invariants can be written in this case. In consequence, the CP-odd EW high-energy chiral Lagrangian describing bosonic interactions, up to four derivatives, contains in total six operators:
\beq
\label{LLG}
\LL_{\text{high,{\cancel{CP}}}}=\ct_{W\Ws}\cBt_{W\Ws}+\ct_{B\Sigma^*}\cBt_{B\Sigma^*}+\ct_{W\Sigma^*}\cBt_{W\Sigma^*}+
\ct_{1}\,\cBt_{1}+\ct_{2}\,\cBt_{2}+\ct_{3}\,\cBt_{3}\,,
\eeq
with
\beq
\begin{aligned}
\cBt_{W\Ws}&=-\dfrac{g^2}{4} \Tr\left(\WLts_{\mu \nu}\WLt^{\mu \nu}\right)
\qquad\,\,
&\cBt_{1}  &= g\,g'\,\Tr\left(\WLts_{\mu \nu}\SH\BLt^{\mu \nu}\SH^{-1}\right) \\
\cBt_{B\Sigma^*} &=g^{\prime 2}\Tr\left(\BLts_{\mu \nu}\SH\BLt^{\mu \nu}\SH^{-1}\right)
&\cBt_{2}  &= \epsilon_{\mu\nu\rho\sigma}\Tr\left(\TLt\left[\VLt^\mu\,,\VLt^\nu\right]\right)\Tr\left(\TLt\left[\VLt^\rho\,,\VLt^\sigma\right]\right)\\
\cBt_{W\Sigma^*} &=g^2\Tr\left(\WLts_{\mu \nu}\SH\WLt^{\mu \nu}\SH^{-1}\right)
&\cBt_{3}  &= \epsilon_{\mu\nu\rho\sigma}\Tr\left(\VLt^\mu\,\VLt^\nu\VLt^\rho\,\VLt^\sigma\right)\,.
\end{aligned}
\label{AppelquistBasisG}
\eeq
According to the effective field theory approach, all the coefficients $\ct_i$ are expected to be of the same order of magnitude.

Notice that the first operator listed in Eq.~(\ref{AppelquistBasisG}), $\cBt_{W\Ws}$, is a topological structure, analogous to the QCD $\theta$--term.  Even if this coupling is usually not considered in the SM context, it is included here in the basis for sake of completeness as it could give rise to non-vanishing effects when considering the incorporation of fermions. On the other side, a similar term for the $U(1)_Y$ is identically vanishing.

Notwithstanding $\theta$--terms, $\LL_{\text{high},{\cancel{CP}}}$ contains five independent operators, and thus at most 
five arbitrary coefficients $\tilde{c_i}$ need to be introduced. They will govern the projection of $\LL_{\text{high}, 
{\cancel{CP}}}$ into $\LL_{\text{low},{\cancel{CP}}}$, in addition to $\xi$. Of course these considerations hold only at tree-level. As the gauging of the SM symmetry breaks explicitly the custodial and the grading symmetries, custodial and/or grading symmetry breaking operators will arise once quantum corrections induced by SM interactions are going to be considered. But this is beyond the scope of this paper.

In the next sections, three exemplifying CH models will be considered.  
%
%
\section{Specific composite Higgs models}
\label{Sect:CHmodels}

In this section, we ``decompose'' the high-energy CP-odd operators of $\LL_{\text{high,{\cancel{CP}}}}$ described 
in the previous section in terms of the low-energy operators of the EW effective chiral Lagrangian $\LL_{\text{low,{\cancel{CP}}}}$, 
recalled in Sec.~\ref{Sect:ExtendedALF}. In particular we consider the following three representative CH models: the $SU(5)/SO(5)$ 
Georgi and Kaplan model~\cite{Georgi:1984af}, the $SO(5)/SO(4)$ Minimal (custodial preserving) Composite Higgs Model
\cite{Agashe:2004rs} and the Minimal (custodial breaking) $SU(3)/(SU(2)\times U(1))$ one.

The following procedure consists in ``decomposing'' the high-energy GB field $\SH(x)$ defined in Eq.~(\ref{Sigma}) 
in terms of its SM and BSM scalar components (if present), and then projecting into the SM d.o.f. (the three would be GBs and 
the Higgs particle), assuming that the extra d.o.f. can be safely decoupled from the observed low-energy spectrum.

%
%
\boldmath
\subsection[The \boldmath $SU(5)/SO(5)$ composite Higgs model]{The $SU(5)/SO(5)$ composite Higgs model}
\unboldmath
\label{Sect:GKmodel}

The first CH model was proposed by Georgi and Kaplan~\cite{Georgi:1984af} more than 30 years ago and encodes the spontaneous 
symmetry breaking $SU(5)\to SO(5)$. It is a non-minimal model in terms of d.o.f. as fourteen GBs arise in the breaking: three of them can be identified with the SM GBs, a fourth one with the physical Higgs particle, while the remaining ten represent new scalar states. In Ref.~\cite{Dugan:1984hq}, it was shown that strong dynamical effects can 
induce large (i.e. ${\cal O}(f)$) masses for all the extra GBs, that therefore can be safely disregarded at low-energies, 
while still having a light (i.e. ${\cal O}(v)$) Higgs particle The discussion of this mechanism is beyond the scope of 
this letter. Moreover, in what follows, all the extra GBs are assumed heavy and removed from the low-energy spectrum, that 
contains only the four SM degrees of freedom. 

For a comprehensive review of the notation used one can refer to Ref.~\cite{Alonso:2014wta}. Here we only briefly recall 
the form of the $\SH(x)$ field. We denote by $\cX(x)$ the $SU(5)$ embedding of the SM GBs:
\beq
\cX(x) =
\dfrac{1}{\sqrt2} 
\begin{pmatrix}
 0 & 0 & \UH(x) e_1 \\
 0 & 0 & \UH(x) e_2  \\
 (\UH(x) e_1)^\dagger & (\UH(x) e_2)^\dagger & 0  \\
\end{pmatrix}\,,  
\label{SU5higgs}
\eeq
with $\UH(x)$ defined as in Eq.~(\ref{SMGBs}) and $e_1^T=(1,0),\,e_2^T=(0,1)$. Once the extra GBs are integrated out, at 
energies below $f$ the GB chiral field $\SH(x)$ can be approximated by:
\beq 
\SH(x)\approx e^{i\frac{\varphi(x)}{f}\cX(x)}\,.
\label{ThetaSU5}
\eeq
with $\varphi(x)$ identified as the field driving the EW symmetry breaking and acquiring a non-vanishing vev,
\beq
\frac{\varphi(x)}{f} \equiv \frac{h(x)+\vh}{f} = \left(\frac{h(x)+\vh}{v}\right) \sqrt{\xi}
\label{varphiExp}
\eeq
and $h(x)$ referring to the physical Higgs particle. Thanks to the special form of the $\cX(x)$, one can easily obtain 
the following expression for $\SH(x)$: 
\beq
\SH(x) = \unity + i\sin\left(\frac{\varphi(x)}{f}\right)\,\cX(x)+\left(\cos\left(\frac{\varphi(x)}{f}\right)-1\right) \cX^2(x)\,.
\label{ThetaSU5bis}
\eeq

The last ingredient that needs to be specified is the embedding of the $SU(2)_L\times U(1)_Y$ generators in $SU(5)$: 
\beq
Q^a_L=\dfrac{1}{2}\left(
\begin{array}{ccc}
 \sigma_a &  &  \\
  & \sigma_a &  \\
  &  & 0 \\
\end{array}
\right)\,,\qquad\qquad
Q_Y=\dfrac{1}{2}\left(
\begin{array}{ccc}
 -\unity_2 &  &  \\
  & \unity_2 &  \\
  &  & 0 \\
\end{array}
\right)\,,
\label{Qa}
\eeq
where $\sigma_a$ denote the Pauli matrices and the normalisation of the generators has been chosen such that $\Tr(Q_a Q_a)=1$. 

The global $SU(2)_L\times SU(2)_R$ symmetry can be embedded in the residual $SO(5)$ group and this implies an approximate 
custodial symmetry conservation.

\subsubsection{The low-energy effective EW chiral Lagrangian}

Having the  explicit expressions for $\SH(x)$, $\VLt_\mu$, $\WLt_\mu$ and $\BLt_\mu$ and substituting them in the 
operators of the high-energy basis of Eq.~(\ref{AppelquistBasisG}) one derives the low-energy effective Lagrangian, 
$\LL_\text{low,{\cancel{CP}}}$ for the Georgi-Kaplan model as a function of the SM would-be GBs, the light scalar field 
$\varphi(x)$ and the SM gauge bosons.

The low-energy projection of the four-derivative effective operators of Eq.~(\ref{AppelquistBasisG}) gives:
\beq
\begin{aligned} 
\cBt_{W\Ws}=&\, \cS_{W\Ws} \,, \\
\cBt_{B\Sigma^*}=&\,-4\,\cS_{B\Bs} +4\,\sin^2\alfm\cS_{B\Bs}  \,, \\
\cBt_{W\Sigma^*}=&\,-4\,\cS_{W\Ws} + 4\,\sin^2\alfm\cS_{W\Ws}\,,\\
\cBt_1 =&\,\frac{1}{2} \sin^2\alfm\cS_1 \,, \\
\cBt_2 =&4\sin^4\alfm\left(\cS_{B\Bs}-\cS_{W\Ws}\right)+\,2\sqrt\xi\,\cos\alfm\sin^3\alfm\left(\cS_2+2\cS_3\right) \,, \\
\cBt_3 =&\,0 \,.
\label{CPLSU5n4}
\end{aligned}
\eeq

The Higgs-independent part of the operator $\cBt_{B\Sigma^*}$ can be safely neglected at low-energy being equivalent to a total derivative and vanishing. On the other side, the Higgs-independent part of the operators $\cBt_{W\Ws}$
and $\cBt_{W\Sigma^*}$ does not vanish as it provides CP-odd non-perturbative contributions. The last operator in the list, 
$\cBt_3$ automatically vanishes in $SU(5)/SO(5)$ model due to the specific properties of the coset generators. Consequently 
at low-energy only four independent CP-odd perturbative couplings are relevant: the Higgs-dependent parts contained in 
$\cBt_{B\Sigma^*}$, $\cBt_{W\Sigma^*}$, $\cBt_1$ and $\cBt_2$.

%
%

\boldmath
\subsection[The minimal \boldmath$SO(5)/SO(4)$ composite Higgs model]{The minimal $SO(5)/SO(4)$ composite Higgs model}
\unboldmath
\label{Sect:MinimalSO5}
A very well know CH model is based on the coset $SO(5)/SO(4)$ \cite{Agashe:2004rs}, with only four GBs arising from the 
global symmetry breaking, that can be identified to the SM GBs and the Higgs field. Moreover as the preserved group $SO(4)$ 
contains $SU(2)_L \times SU(2)_R$, custodial violating effects can arise only through SM-like sources, i.e. proportional 
to the hypercharge (or fermion couplings when introduced).

Due to the minimality of the GB content $\SH(x)$ exactly reduces to 
\beq
\SH(x) = e^{i\frac{\varphi(x)}{f}\cX(x)}\,.
\label{OmegaThetaSO5}
\eeq
with the GB non-linear field given by
\beq
\cX(x) = -\dfrac{i}{\sqrt{2}}\, \Tr\left(\UH \sigma_{\hat{a}}\right) X_{\hat{a}} \,, \qquad \qquad \hat{a}=1,\dots,4\,,
\eeq 
where $\sigma_{\hat{a}}\equiv\left\{\sigma_1,\sigma_2,\sigma_3,i\unity_2\right\}$ and the $SO(5)/SO(4)$ generators 
can be written in a compact form as:
\beq
\left(X_{\hat a}\right)_{ij}=\frac{i}{\sqrt{2}}\left(\delta_{i 5}\delta_{j \hat a}-\delta_{j 5}\delta_{i \hat a}\right)\,,\qquad\qquad 
\hat a=1,\dots,4\,,
\eeq
Alike to the case of the Georgi-Kaplan model, the field $\SH(x)$ takes the simple form in terms of linear and quadratic powers of 
{${\mathcal X}(x)$ shown in Eq.~(\ref{ThetaSU5bis}).

Finally, the embedding of the $SU(2)_L\times U(1)_Y$ generators in $SO(5)$ reads: 
\beq
\begin{aligned}
Q^1_L&=\dfrac{1}{2}\left(
\begin{array}{ccc}
  & -i \sigma_1 &  \\
  i \sigma_1 &  &  \\
  &  & 0 \\
\end{array} \right)\,,\qquad 
& Q^2_L&=\dfrac{1}{2}\left(
\begin{array}{ccc}
  & i \sigma_3 &  \\
  -i \sigma_3 & &  \\
  &  & 0 \\
\end{array} \right)\,,\\
Q^3_L&=\dfrac{1}{2}\left(
\begin{array}{ccc}
 \sigma_2 &  &  \\
  & \sigma_2 &  \\
  &  & 0 \\
\end{array}\right)\,,
\qquad
&Q_Y&=\dfrac{1}{2}\left(
\begin{array}{ccc}
 \sigma_2 &  &  \\
  & -\sigma_2 &  \\
  &  & 0 \\
\end{array} \right)\,.
\end{aligned}
\eeq

\subsubsection{The low-energy effective EW chiral Lagrangian}

The $\LL_\text{low,{\cancel{CP}}}$ Lagrangian originated from the projection of the basis in Eq.~(\ref{AppelquistBasisG}) 
at low energies for the $SO(5)/SO(4)$ minimal CH setup turns out to be the same as that for the $SU(5)/SO(5)$ model. 
Indeed, on one side, the extra GBs present only in the $SU(5)/SO(5)$ model are heavy and therefore the spectrum content 
is the same for both models;  on the other side, the gauging of only the SM group washes out the differences between 
the two preserved subgroups: a unitary transformation can be found linking the GB matrices of the two setups, showing 
that they are equivalent at low-energy. As already discussed in Ref.~\cite{Alonso:2014wta} for the CP-even case, this 
suggests that the low-energy effective chiral Lagrangian for a minimal number of GBs, that can be arranged in a doublet of $SU(2)_L$, and approximate custodial symmetry, is the same regardless of the specific ultraviolet completion. 

%
%
\boldmath
\subsection[The \boldmath$SU(3)/(SU(2)\times U(1))$ composite Higgs model]{The $SU(3)/(SU(2)\times U(1))$ composite Higgs model}
\unboldmath
\label{Sect:MinimalSU3}

The last setup considered is the $SU(3)/(SU(2)\times U(1))$ CH model. This setup is minimal, as it contains only four GBs, 
but contrary to the models previously considered, the custodial $SO(4)$ is not contained in the preserved group $\cH$, and 
therefore there is no (approximate) custodial symmetry. One should be aware that this leads to a large tree-level 
contribution to the $T$ parameter, that therefore requires some level of fine-tuning on the parameter $\xi$, making this 
setup somehow less attractive. Nevertheless, the study of the low-energy projection is instructive: although in the 
initial high-energy $SU(3)/(SU(2)\times U(1))$ Lagrangian no extra sources of custodial breaking besides the SM (gauge) 
one are present, through the decomposition procedure, custodial breaking effects are generated in the low-energy 
effective Lagrangian.

Due to the minimality of the GB content the $\SH(x)$ exactly reduces to 
\beq
\SH(x) = e^{i\frac{\varphi(x)}{f}\cX(x)}\,,
\label{OmegaThetaSO5}
\eeq
with the GB non-linear field given by
\beq
\cX(x) = 
\begin{pmatrix}
 0 & \UH(x) e_2  \\
 (\UH(x) e_2)^\dagger & 0  \\
\end{pmatrix}\,. 
\eeq 
As for the two  models previously analysed, the GB field matrix $\SH(x)$ can be expressed in terms of $\cX(x)$ 
as in Eq.~(\ref{ThetaSU5bis}). Finally the $SU(3)$-embedding of the $SU(2)_L\times U(1)_Y$ generators are given by
\beq
Q^a_L=\dfrac{1}{2}\left(
\begin{array}{cc}
 \sigma_a &    \\
  &  0 \\
\end{array}
\right),
\qquad\qquad
Q_Y=\dfrac{1}{6}\left(
\begin{array}{cc}
 \unity_2 &    \\
  &  -2 \\
\end{array}
\right)\,,
\eeq
with\footnote{A typo is present in Ref.~\cite{Alonso:2014wta} after Eq.~(6.6): $\tr(Q^a_L Q^a_L)= 1/2$ has been adopted also there.} $\tr(Q^a_L Q^a_L)= 1/2$ and $\tr(Q_Y Q_Y)= 1/6$.

\boldmath
\subsubsection{The low-energy effective EW chiral Lagrangian}
\unboldmath 

The low-energy Lagrangian $\LL_\text{low,{\cancel{CP}}}$ is obtained by substituting the explicit expressions for 
$\SH(x)$, $\VLt_\mu$, $\WLt_\mu$ and $\BLt_\mu$ in the operators of the high-energy basis in Eq.~(\ref{AppelquistBasisG}). 
The low-energy projection of the four-derivative operators  listed in Eq.~(\ref{AppelquistBasisG}) results in the following 
decomposition for the $SU(3)/(SU(2)\times U(1))$ model\footnote{This projection is consistent with the one in Ref.~\cite{Alonso:2014wta} in Eq.~(6.8), where a typo is present on the first two operators: the correct values are $\cAt_B=\cP_B/6$, and $\cAt_W=\cP_W/2$.}:
\beq
\begin{aligned} 
\cBt_{W\Ws}=&\,\dfrac{1}{2}\cS_{W\Ws} \,, \\
\cBt_{B\Sigma^*}=&\,-\dfrac{1}{6}\left(1+3 \cos\alfd\right)\cS_{B\Bs}  \,, \\
\cBt_{W\Sigma^*}=&\,-2 \cos\alf\cS_{W\Ws} +\dfrac{1}{2}\sin^4\alfm\cS_{8}\,,\\
\cBt_1 =&\,\frac{1}{8} \sin^2\alf\cS_1 \,, \\
\cBt_2 =&\,\dfrac{1}{4}\sin^4\alf\left(\cS_{B\Bs}-\cS_{W\Ws}\right)+\,\dfrac{1}{4}\sqrt\xi\,\cos\alf\sin^3\alf\left(\cS_2+2\cS_3\right) \,, \\
\cBt_3 =&\,0 \,.
\label{CPLSU3n4}
\end{aligned}
\eeq
Notice, in particular, the presence of the custodial violating operator $\cS_8$ in the decomposition
of the $\cBt_{W\Sigma^*}$ operator, that corresponds to a tree-level source of custodial symmetry breaking. Notice however that it does not contribute to the $T$ parameter and therefore no constraint can be put on its coefficient.

%
\section{Matching the high- and the low-energy Lagrangians}
\label{Sect:Matching}

The functions, appearing in the low-energy basis in each of the CH models considered, encode the 
dependence on the $h$ field: they turn out to be trigonometric due to the GB nature of the Higgs field in these setups. In the general $\LL_\text{low,{\cancel{CP}}}$ basis, this dependence is encoded into the generic functions $\cF_i(h)$ in 
Eq.~(\ref{LALF}) and into some operators which contain derivatives of $h$. It is then possible to identify 
the products $c_i\,\cF_i(h)$ in terms of the high-energy parameters, by comparing the low-energy EW chiral 
Lagrangian of the specific CH models and the general $\LL_\text{low,{\cancel{CP}}}$. This is useful to point out 
specific correlations between couplings that could help investigating the nature of the EWSB mechanism~
\cite{Brivio:2013pma,Brivio:2014pfa,Gavela:2014vra}.

Table \ref{tableProjection} reports the expression of the products $c_i\,\cF_i(h)$ for the three distinct 
CH setups considered before, only for the operators of the low-energy basis that indeed receive contributions.

\begin{table}[h!]
\centering
{\footnotesize
\renewcommand{\arraystretch}{2}
\hspace*{-1.4cm}

\begin{tabular}{|>{$}c<{$}|*2{>{$}c<{$}}|}
\hline
c_i\cF_i(h)& \parbox{2cm}{$SU(5)/SO(5)$\\$SO(5)/SO(4)$} & SU(3)/(SU(2)\times U(1)) \\[1.7mm] 
\hline
\cF_{B\Bs}(h)
&4\,\ct_{B\Sigma^*}\sin^2\frac{\varphi}{2f} + 4\,\ct_{2}\sin^4\frac{\varphi}{2f}
&4\,\ct_{B\Sigma^*}\sin^2\frac{\varphi}{2f}\cos^2\frac{\varphi}{2f} + 4\,\ct_{2}\sin^4\frac{\varphi}{2f}\cos^4\frac{\varphi}{2f}
\\
\cF_{W\Ws}(h)
&\ct_{WW*}-4\,\ct_{W\Sigma^*}\cos^2\frac{\varphi}{2f}- 4\,\ct_{2}\sin^4\frac{\varphi}{2f}
&\frac{\ct_{WW*}}{2}-2\ct_{W\Sigma^*}(1+2\cos^2\frac{\varphi}{2f})- 4\,\ct_{2}\sin^4\frac{\varphi}{2f}\cos^4\frac{\varphi}{2f}

\\
c_1\cF_1(h)
&\dfrac{\ct_1}{2}\sin^2\frac{\varphi}{2f}
&\dfrac{\ct_1}{2}\sin^2\frac{\varphi}{2f}\cos^2\frac{\varphi}{2f}
\\
c_2\cF_2(h)
& 2\,\ct_{2}\sqrt\xi\cos\frac{\varphi}{2f}\sin^3\frac{\varphi}{2f}
& 2\,\ct_{2}\sqrt\xi\cos^3\frac{\varphi}{2f}\sin^3\frac{\varphi}{2f}\left(2\cos^2\frac{\varphi}{2f}-1\right)
\\
c_3\cF_3(h)
&4\,\ct_{2}\sqrt\xi\cos\frac{\varphi}{2f}\sin^3\frac{\varphi}{2f}
& 4\,\ct_{2}\sqrt\xi\cos^3\frac{\varphi}{2f}\sin^3\frac{\varphi}{2f}\left(2\cos^2\frac{\varphi}{2f}-1\right)
\\[1mm]
\hline
c_8\cF_8(h)
&-
&\dfrac{\ct_{W\Sigma^*}}{2}\sin^4\frac{\varphi}{2f}
\\[1mm]
\hline
\end{tabular}}
\caption{\em Expressions for the products $c_i\,\cF_i(h)$ for $SU(5)/SO(5)$ ($SO(5)/SO(4)$) and $SU(3)/(SU(2)\times U(1)$ 
respectively. The ``$-$'' entry indicates no leading order contributions at low-energy to the corresponding operator.}
\label{tableProjection}
\end{table}

Some interesting comments can be inferred from Tab.~\ref{tableProjection}.
\begin{description}
\item[\boldmath $SU(5)/SO(5)$ and $SO(5)/SO(4)$ --] All the custodial preserving operators entering the low-energy 
Lagrangian $\LL_\text{low,{\cancel{CP}}}$ result from the projection. As expected, no tree-level custodial breaking 
operator arises from the low-energy projection of $SU(5)/SO(5)$ and $SO(5)/SO(4)$ models, as they are naturally 
custodial preserving. Moreover, notice that $c_2\cF_2(h)$ and $c_3\cF_3(h)$ are functions of the same high-energy 
parameter $\ct_2$, implying a correlation between the couplings described by $\cS_2$ and $\cS_3$.

\item[\boldmath $SU(3)/(SU(2)\times U(1))$ --] Besides the custodial preserving operators, only one custodial breaking 
operator of the low-energy basis, $\cS_8$, receives contributions from the projection. As for the previous CH models, 
the interactions described by $\cS_2$ and $\cS_3$ turn out to be correlated.
\end{description}
Furthermore, notice that the arbitrary functions $\cF_i(h)$ of the generic low-energy effective chiral Lagrangian 
become now a constrained set, as a consequence of having chosen a specific CH model.
 
 \boldmath
 \subsection{The small $\xi$ limit}
 \unboldmath
 
It is particularly interesting to consider the $f\gg v$, or equivalently $\xi\ll1$, limit. In fact in this limit the 
non-linear CH model should overlap with the case in which the EWSB is linearly realised and the Lagrangian written 
in terms of the Higgs as an $SU(2)_L$ doublet. For example, taking the operators in Eq.~(\ref{CPLSU5n4}) for the $SU(5)/SO(5)$ setup and expanding them in Taylor series in $1/f$ as defined in
Eq.~(\ref{varphiExp}), one concludes that at first order in $\xi$,
\beq
\begin{aligned}
\cBt_{B\Sigma^*}&\approx \xi\,(1+h/v)^2\cS_{B\Bs}-4\, \cS_{B\Bs}\\
\cBt_{W\Sigma^*}&\approx \xi\,(1+h/v)^2\cS_{W\Ws}-4\, \cS_{W\Ws}\\
\cBt_1 &\approx\frac{1}{8}\,\xi\,(1+h/v)^2\cS_1 \\
\cBt_2 &\approx 0 \,
\end{aligned}
\eeq
with $\cBt_2$ giving contribution only at order $O(\xi^2)$. This procedure can be repeated with all the terms appearing in Table~\ref{tableProjection} and the results can be read in Table~\ref{tableProjection2}.
\begin{table}[h!]
\centering
{\footnotesize
\renewcommand{\arraystretch}{2}
\begin{tabular}{|>{$}c<{$}|*2{>{$}c<{$}}|}
\hline
c_i\cF_i(h)& \parbox{2cm}{$SU(5)/SO(5)$\\$SO(5)/SO(4)$} & SU(3)/(SU(2)\times U(1)) \\[1.7mm] 
\hline
\cF_{B\Bs}(h)
&\ct_{B\Sigma^*}\, \xi \left(1+\dfrac{h}{v}\right)^2 + \mathcal{O}(\xi^2)
&\ct_{B\Sigma^*}\, \xi \left(1+\dfrac{h}{v}\right)^2 + \mathcal{O}(\xi^2)
\\
\cF_{W\Ws}(h)
&\ct_{WW*}+\ct_{W\Sigma^*} \left(-4+\xi \left(1+\dfrac{h}{v}\right)^2\right)+ \mathcal{O}(\xi^2)
&\dfrac{\ct_{WW*}}{2}+\ct_{W\Sigma^*}\left(-6+\xi \left(1+\dfrac{h}{v}\right)^2\right) + \mathcal{O}(\xi^2)
\\
c_1\cF_1(h)
&\dfrac{\ct_1}{8}\, \xi \left(1+\dfrac{h}{v}\right)^2 + \mathcal{O}(\xi^2)
&\dfrac{\ct_1}{8}\, \xi \left(1+\dfrac{h}{v}\right)^2 + \mathcal{O}(\xi^2)
\\
c_2\cF_2(h)
& \mathcal{O}(\xi^2)
& \mathcal{O}(\xi^2)
\\
c_3\cF_3(h)
& \mathcal{O}(\xi^2)
& \mathcal{O}(\xi^2)
\\[1mm]
\hline
c_8\cF_8(h)
&-
& \mathcal{O}(\xi^2)
\\[1mm]
\hline
\end{tabular}}
\caption{\em Expressions for the products $c_i\,\cF_i(h)$ for $SU(5)/SO(5)$ ($SO(5)/SO(4)$) and $SU(3)/(SU(2)\times U(1))$ in the $\xi \ll 1$ limit.}
\label{tableProjection2}
\end{table}
This should be compared with the effective $d=6$ CP-odd Lagrangian in the linear regime~\cite{Buchmuller:1985jz,Grzadkowski:2010es}: only three operators form the 
EW bosonic basis:
\beq
\begin{aligned}
\cQ_{\varphi\widetilde B}&=\BLs_{\mu\nu}\BL^{\mu\nu} \Phi^\dag\Phi\\
\cQ_{\varphi\widetilde W}&=\Phi^\dag\WLs_{\mu\nu}\WL^{\mu\nu} \Phi\\
\cQ_{\varphi\widetilde BW}&=\BLs_{\mu\nu} \Phi^\dag\WL^{\mu\nu}\Phi\,,
\end{aligned}
\eeq
and own physical interactions with perturbative effects. There is one-to-one correspondence between these 
two classes of operators: $\cQ_{\varphi\widetilde B}\leftrightarrow \cBt_{B\Sigma^*}$, $\cQ_{\varphi\widetilde W}
\leftrightarrow \cBt_{W\Sigma^*}$ and $\cQ_{\varphi\widetilde BW}\leftrightarrow \cBt_1$. Conversely, $\cBt_2$ 
contributes at low-energy to $\cS_2$ and $\cS_3$, but only at $\xi^2$, i.e. its linear sibling\footnote{This is in 
contrast with Eq.~(A.1) in Ref.~\cite{Gavela:2014vra}, where two $d=6$ linear operators have been indicated 
as siblings of $\cS_2$ and $\cS_3$. Those operators do contain the interactions of $\cS_2$ and $\cS_3$, but they 
are not the lowest dimensional ones.} should have $d=8$: indeed, by using integration by parts and the Bianchi identities, 
it is straightforward to verify that the interactions of  $\cS_2$ and $\cS_3$ are described at the lowest order in the 
linear expansion by the operators 
\beq
\BLs_{\mu\nu}\left(\Phi^\dag \Dfb \Phi\right) \DL^\nu \left(\Phi^\dag \Phi\right)\,,\qquad
\left(\Phi^\dag \Dfb \WLs_{\mu\nu}\Phi\right) \DL^\nu \left(\Phi^\dag \Phi\right)
\eeq
with  $\DL_\mu\Phi\equiv \left(\partial_\mu+ \frac{i}{2} g' B_\mu + \frac{i}{2}g\sigma_i W^i_\mu \right)\Phi $ and 
$\Phi^\dag\Dfb\Phi\equiv\Phi^\dag\DL_\mu\Phi-\DL_\mu\Phi^\dag\Phi$.

The products $c_i\cF_i(h)$ corresponding to custodial-breaking operators are suppressed by $\xi^2$ and therefore 
they are also described in the linear expansion by $d=8$ operators. However, a complete comparison is not possible 
in this case, as no $d=8$ basis has been defined yet, to our knowledge.

%
%
\section{Conclusions}
\label{Sect:Conc}

The CP-odd Lagrangian for a generic $\cG/\cH$ symmetric coset has been constructed, focusing on the bosonic sector: only 
four operators constitute a basis. Once considering the gauging of the hypercharge as the only source of custodial 
symmetry breaking, the number of operators increases up to six, including the topological term. These operators 
are written in terms of the SM gauge bosons, embedded in $\cG$ representations, and of the matrix for the Goldstone 
Bosons, which in these contexts also includes the physical Higgs fields. Once a specific coset is considered, however, 
some of these operators could vanish or be redundant.

The projection of this Lagrangian at low-energy for three different CH models has allowed to confirm the subset of custodial preserving couplings of the general 
low-energy CP-odd Lagrangian for a dynamical Higgs developed in Ref.~\cite{Gavela:2014vra} and leads to the identification 
of correlations among the effective low-energy operators and allows a comparison with the linear effective Lagrangian 
for an elementary Higgs. The results presented here are consistent with Ref.~\cite{Alonso:2014wta}, where only the 
CP-even sector has been considered. In particular, the low-energy projection of the $SU(5)/SO(5)$ and $SO(5)/SO(4)$ 
models turns out to be the same, due to the fact that neglecting the effects of the extra GBs of the first model and 
gauging only the SM group force the preserved subgroups in the two models to be isomorphic. On the other hand, some 
differences appear with the low-energy projection of $SU(3)/(SU(2)\times U(1))$, that is not intrinsically custodial 
preserving.

The results also confirm the powers of $\xi$ predicted in Ref.~\cite{Gavela:2014vra} as  weights for each operator 
of the low-energy effective chiral Lagrangian, allowing an immediate comparison with linear expansions for an elementary Higgs. One can point out that the differences stem from the $h$ dependence: functions of $\sin\left[(\vh+h)/2f\right]$ for the CH models and powers of $\left(v+h\right)/2$ for the linear realisation. When $\xi\ll1$, the trigonometric dependence on $h$ reduces exactly to the linear one, as $\sin^2(\varphi/f) = \xi(1+h/v)^2+\cO(\xi^2)$, neglecting the higher order terms in $\xi$. This result suggests that the use of the linear expansion to construct CH model Lagrangians can be justified in this limit.  On the other hand, if $\xi$ is not so small, the deviations from the linear structure $(1+h/v)$ could be significant and therefore comparing observables with different Higgs legs could disentangle an elementary from a composite Higgs scenario. 
 
Finally, to distinguish between the doublet or non-doublet nature of the Higgs, it is then necessary to compare 
pure-gauge and gauge-Higgs couplings~\cite{Brivio:2013pma,Brivio:2014pfa,Gavela:2014vra}, whose precise form we have 
determined here for the specific CH models considered. The strength of these observables depends on $\xi$ and therefore 
the larger $\xi$ the easier it will be to detect a signal that may be able to shed light on the Higgs representation.

\section*{Acknowledgements}
We thank Ilaria Brivio, Roberto Contino, Belen Gavela, Elizabeth Jenkins, Aneesh Manohar and Andrea Wulzer for 
constructive conversations. The work of I.M.H. is supported by an ESR contract of the European Union network 
FP7 ITN INVISIBLES (Marie Curie Actions, PITN-GA-2011-289442). L.M. acknowledges partial support of the 
European Union network FP7 ITN INVISIBLES, of CiCYT through the project FPA2012-31880, of the Spanish MINECO's ``Centro de Excelencia Severo Ochoa'' Programme under grant SEV-2012-0249, and by a grant from the Simons Foundation. S.R. acknowledges partial support of the European Union network FP7 ITN INVISIBLES and of the COFIN program PRIN 2010. This work was partially performed at the Aspen Center for Physics, which is supported by National Science Foundation grant PHY-1066293.



\providecommand{\href}[2]{#2}\begingroup\raggedright\endgroup

\end{document}